\documentclass[12pt]{article}
\newcommand{\CM}{{\mathrm CM}}

\title{Event generation of large-angle Bhabha scattering 
at LEP2 energies}

\author{A.B. Arbuzov\thanks{on leave of absence from 
Joint Institute for Nuclear Research, Dubna, Russia.}
}

\date{}

\begin{document}
\maketitle

\begin{itemize}
\item[$ $]
        {\em Dipartimento di Fisica Teorica, Universit\`a di Torino, \\
             INFN, Sezione di Torino, \\
             via Giuria 1, I-10125 Torino, Italy \/} \\
        {\tt e-mail: arbuzov@to.infn.it}
\end{itemize}

\begin{abstract}
LABSMC Monte Carlo event generator is used to simulate Bhabha
scattering at high energies. Different sources of radiative 
corrections are considered. The resulting precision is discussed.
\\[.2cm] \noindent
\end{abstract}

\section{Introduction}

In this note we are going to discuss 
the application the Monte Carlo event generator 
{\tt LABSMC}~\cite{LABSMC} to the case of LEP2 energies.
The topic is actual now in view of the analysis
of LEP2 data. The experiment requires high precision theoretical
predictions to perform more deep tests of the Standard
Model and to look for a new physics.

Initially the {\tt LABSMC} event generator was developed 
to simulate large--angle Bhabha scattering at energies
of about a few GeV's at electron positron colliders like
VEPP--2M and DA$\Phi$NE. The code included the
Born level matrix element, the complete set of ${\cal O}(\alpha)$ 
QED RC, and the higher order leading logarithmic RC by means
of the electron structure functions. The relevant set of formulae
can be found in Ref.~\cite{JHEP97a}. The generation of events is
performed using an original algorithm, which combines advantages of
semi--analytical programs and Monte Carlo generators.

The structure of our event generator was described recently in 
paper~\cite{LABSMC}. The extension for higher energies is done
by introducing electroweak (EW) contributions, such as $Z$-exchange, 
into the matrix elements. The third~\cite{Skr} and fourth~\cite{p4}
order leading logarithmic
photonic corrections were also included in the new version.
So, the structure of the code is kept the same, and we have to describe
now what kind of EW effects were included in our code to work
with large--angle Bhabha
at LEP2. In particular we are going to consider the region
of radiative return to the $Z$-peak in radiative Bhabha scattering.
To estimate the resulting theoretical uncertainty, one has 
to analyse various sources of radiative corrections (RC).

\section{Electroweak contributions}

The set of electroweak effects 
was included according to Refs.~\cite{BDH,BBM,BKH}. 
The Born level cross section in the program contains
now both the photon and $Z$-boson exchange contributions (it is used
also as a kernel cross section for higher order leading log radiative
corrections).
The first order virtual and soft EW RC were taken directly from the
semi--analytical code {\tt ALIBABA}~\cite{BBM}. 
The EW matrix element for radiative Bhabha is taken from Ref.~\cite{BKH}. 
A comparison of the Bhabha cross--section
integrated over photons is in a reasonably good agreement  
(see Table~1, error--bars are dropped)
with the published numbers of other codes~\cite{YR9601}.

\begin{table}[ht]
\caption{Comparison with Fig.~21 from Ref.~\cite{YR9601},
cross-sections in pb.}
\begin{tabular}[]{|c|c|c|c|c|c|c|c|}
\hline
$E_{\CM}$, GeV & {\tt BHWIDE} & {\tt TOPAZ0} & {\tt BHAGENE3} &
{\tt UNIBAB} & {\tt SABSPV} & {\tt BHAGEN95} & {\tt LABSMC} \\ \hline
& \multicolumn{7}{c|}{$\vartheta_{\mathrm{acol}}=10^{\circ}$} \\ \hline
175   & 35.257 & 35.455 & 34.690 & 34.498 & 35.740 & 35.847 & 35.337 \\ 
190   & 29.899 & 30.024 & 28.780 & 29.189 & 30.270 & 30.352 & 29.941 \\ 
205   & 25.593 & 25.738 & 24.690 & 25.976 & 25.960 & 26.007 & 25.687 \\ 
\hline
& \multicolumn{7}{c|}{$\vartheta_{\mathrm{acol}}=25^{\circ}$} \\ \hline
175   & 39.741 & 40.487 & 39.170 & 39.521 & 40.240 & 40.505 & 40.029 \\ 
190   & 33.698 & 34.336 & 32.400 & 33.512 & 34.100 & 34.331 & 33.954 \\ 
205   & 28.929 & 29.460 & 27.840 & 28.710 & 29.280 & 29.437 & 29.178 \\ 
\hline
\end{tabular}
\end{table}

\section{Radiative return with a visible photon}

At LEP2 the radiative return to the
$Z$-peak due to photon or pair radiation gives a sizable contribution 
to the cross--section. This process is
used itself in particular
to look for anomalous gauge boson couplings.

The pure tree level matrix element~\cite{BKH} for radiative process
\begin{equation} \label{proc}
e^+\ +\ e^-\ \longrightarrow \ e^+\ +\ e^-\ +\ \gamma\ +\ (n\gamma)
\end{equation}
was supplemented by radiative
corrections due to initial state soft and hard collinear radiation
by means of the electron structure function approach~\cite{SF}.
The electron--positron pair 
production was taken into account in the same way. 
As an energy scale for the structure functions we took the 
$t$-channel momentum transferred, because the corresponding
diagrams are dominant.
The vacuum 
polarization correction to the photon propagators is applied as well.

In Table~2 we put the result for the following conditions:
$E_{\CM}$=183, 189~GeV; $|\cos\theta_{e^\pm}|<0.95$; at least one
electron has $|\cos\theta_{e}|<0.7$; electrons should have transverse
momenta above 1~GeV; the final particles are
to be isolated by at least 20 degree from each other; the total
observed energy $>0.8\ E_{\CM}$; $|\cos\theta_{\gamma}|<0.7$.
In the columns {\it without $Z$-peak} we excluded the events with 
the invariant mass of the electron positron pair in the range 
85~GeV~$<M_{ee}<$~95~GeV.
As could be seen from the numbers, the ISR LLA corrections 
are in this case of the order 2\%. 
The additional non--standard LLA corrections, which were found in
Ref.~\cite{viol}, make a small shift of the correction; but an
independent verification of the investigation is required.

The complete set of ${\mathcal O}(\alpha)$ EW radiative corrections
to the process~(\ref{proc}) is unknown. To estimate the uncertainty
of our result we look at the relative size of the known leading and
sub--leading ${\mathcal O}(\alpha)$ RC to the Bhabha process itself.
For an analogous set of cuts for Bhabha scattering for the difference
of the correction values we have
$\delta_{\mathrm{tot}}-\delta_{\mathrm{LLA}}
\approx 1\%$\footnote{The special cut
on the scattering angle of ``at least one electron has 
$|\cos\theta_e|<0.7$''
is similar to the narrow--wide event selection in small--angle Bhabha
at LEP1. In both cases we see a considerable reduction of the RC size.
If we apply this cut, the difference 
$\delta_{\mathrm{tot}}-\delta_{\mathrm{LLA}}$ is
well below the 1\% 
level.}. 
In this way we estimate 
the precision of our results for the radiative process~(\ref{proc})
to be of the order 1.5\%.

\begin{table}[ht]
\caption{The cross section in pb of radiative Bhabha with a visible
photon in different approximations.}
\begin{tabular}[]{|l|c|c|c|c|}
\hline
                   & \multicolumn{2}{c|}{total} 
& \multicolumn{2}{c|}{without $Z$-peak} \\ \hline
$E_{\CM}$~[GeV]    & 183    & 189    & 183    & 189    \\ \hline  
tree--level        & 0.9817 & 0.9146 & 0.8251 & 0.7727 \\ 
vacuum polarisation& 1.1022 & 1.0342 & 0.9630 & 0.8853 \\
vac. pol. + ISR LLA& 1.0842 & 1.0088 & 0.9346 & 0.8770 \\
\hline
\end{tabular}
\end{table}

\section{Conclusions}

The precision of the theoretical predictions, which can be
obtained by means of the presented code for 
inclusive large--angle Bhabha scattering at LEP2, is estimated
to be of the order 0.2\%. 
It is defined mainly by the unknown radiative corrections of the order 
${\mathcal O}((\alpha/\pi)^2L)\approx 10^{-4}$. The coefficients
before these terms are not too large, as was seen in the case
of small--angle Bhabha scattering. At LEP2 energies in the 
large--angle Bhabha process the $t$-channel photon exchange
is dominant. Nevertheless, a correct treatment of the electroweak
Born and the first order corrections is important.
The technical precision of the code is to be verified in
further test and comparisons with other codes. 
By means of the comparisons of the semi--analytical branch and 
the pure Monte Carlo one of the code we have a good control of
such parameters as the precision of numerical integration
and the number of events to be generated. That allows to reach
an ordered level of the uncertainty in numerical evaluations.

Another source of uncertainties is an incomplete treatment 
of pair production.
In the current version of the code the pair production corrections 
are evaluated in the ${\mathcal O}(\alpha^2L^2)$ structure function
approach. Both the singlet and non--singlet electron pairs are 
included. A special brunch of the code to scrutinise the pair 
production~\cite{labspr} is in progress. 

The inclusion of the third and fourth order LLA photonic corrections
allows not to use exponentiation. A simple estimate shows that the 
difference between the two treatments at LEP2 is negligible,
while the exponentiation requires a quite different event
generation procedure.

\subsection*{Acknowledgements}

I am grateful for support to the INTAS foundation, grant 93--1867 ext.

\end{document}